\documentclass[conference]{IEEEtran}
\IEEEoverridecommandlockouts
\usepackage{amsmath,amsfonts}
\usepackage{array}
\usepackage[caption=false,font=normalsize,labelfont=sf,textfont=sf]{subfig}
\usepackage{textcomp}
\usepackage{stfloats}
\usepackage{amsmath}
\usepackage{eqnarray,amsmath}
\usepackage{url}
\usepackage{amssymb}
\usepackage{lipsum}
\usepackage{cite}
\usepackage[center]{caption}
\usepackage{verbatim}
\usepackage{graphicx}
\usepackage[table]{xcolor}
\usepackage{amsfonts}
\usepackage[left=0.6in,right=0.6in,top=0.64in,bottom=0.97in]{geometry}

\usepackage{hyperref}

\def\BibTeX{{\rm B\kern-.05em{\sc i\kern-.025em b}\kern-.08em
    T\kern-.1667em\lower.7ex\hbox{E}\kern-.125emX}}
\begin{document}

\title{Conference Paper Title*\\
{\footnotesize \textsuperscript{*}Note: Sub-titles are not captured in Xplore and
should not be used}
\thanks{This work was made possible by AICC03-0324-200005 from the Qatar National Research Fund (a member of the Qatar Foundation). The findings herein reflect the work, and are solely the responsibility, of the authors.}
}

\title{A BFF-Based Attention Mechanism for Trajectory Estimation in mmWave MIMO Communications\\}

\author{Mohammad Shamsesalehi\dag , Mahmoud Ahmadian Attari\dag, Mohammad Amin Maleki Sadr\dag\dag, \\Benoit Champagne\ddag, Marwa Qaraqe* \\ \dag Dept. ECE, K. N. Toosi University of Technology, Tehran, Iran, email:\{shamsesalehi,mahmoud\}@email.kntu.ac.ir,   \\
\ddag Dept. ECE, McGill University, Montréal, Canada, email: 
 {benoit.champagne@mcgill.ca}, \\
*College of Science and Engineering, Hamad Bin Khalifa University, Qatar Foundation, Doha, Qatar\\email: *mqaraqe@hbku.edu.qa.  
* aminmalekisadr@gmail.com
\\
\thanks{
This publication was made possible by AICC03-0324-200005 from the Qatar National Research Fund (a member of Qatar Foundation). The findings herein reflect the work, and are solely the responsibility, of the authors.}}
\maketitle

\begin{abstract}
This paper explores a novel Neural Network (NN) architecture suitable for Beamformed Fingerprint (BFF) localization in a millimeter-wave (mmWave) multiple-input multiple-output (MIMO) outdoor system. The mmWave frequency bands have attracted significant attention due to their precise timing measurements, making them appealing for applications demanding accurate device localization and trajectory estimation.
The proposed NN architecture captures BFF sequences originating from various user paths, and through the application of learning mechanisms, subsequently estimates these trajectories. Specifically, we propose a method for trajectory estimation, employing a transformer network (TN) that relies on attention mechanisms. This TN-based approach estimates wireless device trajectories using BFF sequences recorded within a mmWave MIMO outdoor system.
To validate the efficacy of our proposed approach, numerical experiments are conducted using a comprehensive dataset of radio measurements in an outdoor setting, complemented with ray tracing to simulate wireless signal propagation at 28 GHz. The results illustrate that the TN-based trajectory estimator outperforms other methods from the existing literature and possesses the ability to generalize effectively to new trajectories outside the training dataset.

\end{abstract}

 
\section{Introduction}
Beyond 5G networks operating within the millimeter wave (mmWave) frequency bands, will provide unprecedented bandwidth, and are expected to excel in terms of throughput, latency, and reliability across various transmission distances \cite{7413967}.
To achieve these impressive goals, it is required that both transmitting and receiving equipment possess precise knowledge about each other's relative position and orientation \cite{mogyorosi2022positioning}. Moreover, the availability of accurate localization information holds great importance for a range of critical applications, and especially online trajectory estimation \cite{yin2022millimeter,kim20185g}.

From the perspective of the radio environment, localization within mmWave bands can be categorized in terms of indoor and outdoor scenarios. A specific subset of notable studies focused on indoor localization includes the works in \cite{vukmirovic2018position, lin20183, vashist2021kf}.
In \cite{vukmirovic2018position}, a novel massive
MIMO (m-MIMO) framework 
for precise position estimation in the mmWave band is proposed. 
In \cite{lin20183}, the authors introduce a novel hybrid 3D indoor positioning scheme designed for mmWave massive MIMO systems, which combines received signal strength and angle of arrival. 
For the outdoor scenarios, which is the focus of this work, \cite{shen20212d} presents a fingerprinting-based cooperative localization approach tailored for mmWave massive MIMO systems, while \cite{sadr2021uncertainty} introduces a method using Monte Carlo (MC) dropout to capture uncertainty in a CNN-based mmWave MIMO localization system. 
In \cite{mendrzik2019localization}, authors explore the statistical relationship between channel state information (CSI) and specific characteristics of user devices, including position, orientation, and clock offset. 

From a design perspective, localization methods can be further categorized based on their modeling approach, i.e., distinguishing between physical and data-driven methods. 
The former methods rely on physical knowledge of the radio environment and channel characteristics  \cite{koivisto2017joint}. 
However, applying these methods in dense urban environments at mmWave frequencies presents substantial challenges due to the intricate nature of channel models and the multitude of parameters involved \cite{kanhere2018position}.
In contrast, data-driven models adopt a different strategy and estimate user positions by collecting radio data or fingerprints from a broad range of representative locations. Subsequently, machine learning (ML) algorithms are employed to predict the location of a new target based on its radio imprints or fingerprints \cite{8307353}.

In \cite{zhang2017path}, the authors introduce a localization method based on fingerprints that utilize path-loss measurements for both training and prediction phases. Another fingerprint-based method is presented in \cite{zhou2017robust}, where crowd-sourcing is employed to collect probabilistic Radio Signal Strength (RSS) data from numerous smartphones, forming a database of fingerprints.  
Among the fingerprint-based techniques, the Beamformed Fingerprint (BFF) methods stand out. In \cite{gante2018beamformed}, a BFF method is
proposed to achieve precise outdoor positioning by capitalizing on mmWave MIMO transmissions to achieve steerable and highly focused radiation patterns. Recent studies show that BFF methods can generate data-rich information, allowing ML techniques to discern subtle details in both localization and trajectory estimation \cite{gante2020deep}.

While recurrent Neural Networks (RNNs) are a common choice for modeling sequential data, they suffer from gradient-related issues which diminish their effectiveness over time, especially their ability to retain historical information. These issues can be mitigated to some extent by incorporating gating mechanisms into RNN cells, with Long Short-Term Memory (LSTM) being the most prevalent architecture for this purpose. In \cite{bai2018rfedrnn} an LSTM network is successfully employed for mobile localization based on WiFi fingerprints in an indoor setting.
Recently, Transformer Networks (TN), also known simply as transformers, have demonstrated remarkable performance and adaptability in effectively capturing and modeling sequential signals, as outlined in \cite{vaswani2017attention}. However, to the best of our knowledge, TN have not yet been applied to the timely problem of BFF in outdoor MIMO systems.

In this paper, we introduce a deep learning (DL) approach that relies on BFF sequences for trajectory estimation within a mmWave MIMO outdoor system. 
Our approach capitalizes on the unique capabilities of BFF data and employs a specific TN-based architecture to capture long-term historical data for trajectory estimation tasks.
Our main contributions and distinguishing features of our work are summarized as follows: 
\begin{itemize}
\item We develop a novel TN architecture based on attention mechanism
to forecast the trajectories of wireless devices utilizing BFF sequences recorded within a mmWave MIMO outdoor system.
\item Unlike RNN, LSTM, or temporal convolutional networks (TCN),  the use of an encoder-decoder structure along with attention mechanism allows the proposed TN to detect subtle changes in the users’ directions within lengthy sequences, and to generalize effectively to novel trajectories not included in the training set.
\item The TN architecture naturally lends itself to a parallel implementation, thereby enhancing computational speed and processing efficiency across different sequences.
\item We validate the effectiveness of our proposed TN-based approach through numerical experiments, using a substantial dataset of radio measurements collected in an outdoor environment, enhanced by incorporating ray tracing techniques to replicate wireless propagation at 28 GHz frequency.
\item We show that our proposed method can achieve accuracy levels of nearly one meter for moving vehicles and less than one meter for pedestrians, surpassing the performance of benchmark methods from the recent literature.

\end{itemize}
The remainder of this paper is structured as follows: Section II presents the system model for BFF-based localization within a mmWave MIMO outdoor system. Section III introduces the TN architecture based on attention mechanism for addressing the trajectory estimation problem. Section IV presents the simulation methodology along with our key experimental findings. Finally, Section V concludes the paper.
 
\section{System Model and Problem Statement}
As outdoor 5G base stations (BSs) are typically positioned at elevated locations within urban environments, buildings emerge as the primary obstacles that remain stationary over extended periods. Under such conditions, the received power delay profile (PDP) measurements, as detailed in \cite{guan2020channel}, are expected to be a valuable resource for radio localization. To maximize coverage, BSs can employ a sequence of directive Beamforming (BF) patterns to transmit signals, effectively covering all feasible transmission angles. Consequently, receivers can capture multiple distinct PDPs. Because of the intricate propagation mechanisms that occur when obstacles are present, it is expected that the measured PDPs will exhibit significant variations or discontinuities across the localization area. These abrupt variations offer valuable
spatial information to a trained localization model \cite{gante2020deep}.

 
\begin{figure}
\centering
\includegraphics[width=0.4\textwidth]{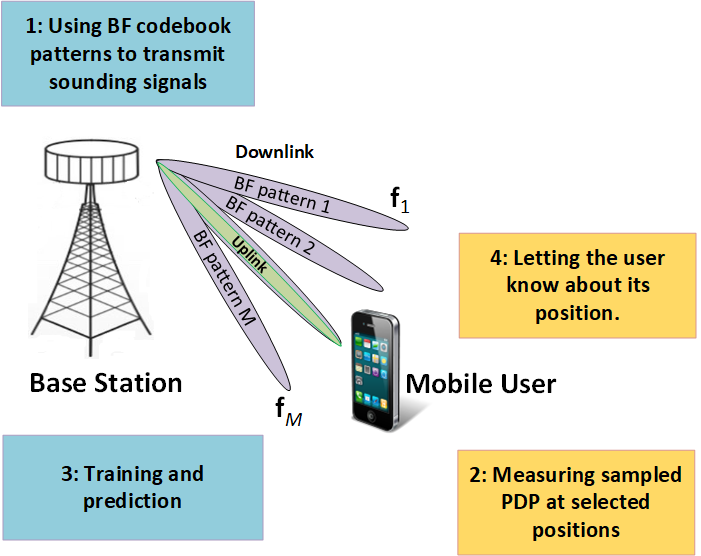}
\caption{\label{fig:frog} System model of BFF-based  MIMO mmWave localization }
\label{system_model}
\end{figure}

The process of obtaining the BFF data through a BF codebook at the BS and PDP monitoring at the mobile user is illustrated in Fig. \ref{system_model}. At the BS, directional sounding signals are generated to facilitate the extraction of BFF data from the radio environment for both purpose of offline model training and online prediction, i.e., trajectory estimation, as elabotated below. 

During the training phase, a BS equipped with $N_T$ antennas transmits signals within a specific frequency band $B$ using a predetermined set of $M$ transmit BF vectors from a codebook $\mathcal{C} = \left\{ \textbf{f}_1, \textbf{f}_2, \cdots , \textbf{f}_M \right\}$, where $\textbf{f}_i\in \mathbb{C}^{N_T\times 1}$ denotes the $i$th BF vector. 
A user terminal equipped with $N_R$ antennas and located at a known position (to be varied) is used to capture the sequence of transmitted signals. The received user signal from the $i$th codebook at a specific sounding frequency bin $f \in B$, can be expressed as:
 \begin{equation}   \label{eq1}
r_i(f)= \textbf{w}^{H} \textbf{H}\textbf{f}_is+\textbf{w}^{H}\textbf{z},\;\;\; i\in \left\{ 1,\cdots ,M \right\},
\end{equation}
where $s\in \mathbb{C}$ is the sounding signal amplitude, $\textbf{H}\in \mathbb{C}^{N_R\times N_T}$ is the complex channel gain matrix between the BS and user device, $\textbf{w}\in \mathbb{C}^{N_R\times 1}$ is the BF vector at the receiver and $\textbf{z}\in \mathbb{C}^{N_R\times 1}$ is an additive noise term. 

At the user terminal, through processing of the received signals $r_i(f)$ over all frequencies $f \in B$, the received power is measured and an overall PDP $P_i(\tau)$ is calculated, where $\tau$ represents the (continuous) delay variable. A discrete-time version of the PDP, denoted as $P_i[j]$, is obtained by uniformly sampling $P_i(\tau)$ at frequency $F_s = {T_s}^{-1}$, where $T_s$ denotes the sampling interval \cite{guan2020channel}. The PDP can be expressed as:
 \begin{equation}   \label{eq2}
P_i[j]= P_i(jT_s),\;\;\; \forall j\in \left\{ 1,\cdots ,N_s \right\}.
\end{equation}
where $N_s = \frac{T }{T_s}$ represents the total number of samples and $T$ denotes the maximum excess delay. The PDP samples are represented by binary values for each sampling time, i.e.,
 \begin{equation}   \label{eq3}
 x_{i,j} = I(P_i[j] \geq \eta),
\end{equation}
where $I$ is the indicator function and $\eta$ is a positive threshold level. By utilizing binary PDP values, implementation complexity and memory requirement of the TN can be decreased. Furthermore, utilizing a binary representation of the PDP can effectively handle high noise levels and improve localization performance \cite{gante2020deep}. Binary PDP values can be used to construct a feature matrix $\textbf{X} = \left[ x_{i,j} \right]$ which is transmitted to the BS for localization purposes.  

During the training phase, PDP data is collected as above for a predetermined grid consisting of user terminal locations. The locations are indexed by $n\in \left\{ 1,\cdots ,\mathcal{N} \right\}$, where $\mathcal{N}$ is the total number of locations, and $\textbf{y}_{n} \in \mathbb{R}^{1 \times 2}$ represents the coordinates of the $n$th location in 2D space. 
For each one of these locations, the PDP feature matrix $\textbf{X}_{n}\in \mathbb{R}^{M\times N_s}$ and the corresponding position label $\textbf{y}_{n}\in \mathbb{R}^{1 \times 2}$ are collected as the input and output data for the TN. These matrices and vectors are collectively referred to as the BFF and are stored in the BS as a dataset $\mathcal{D} =  (\textbf{X}_{n}, \textbf{y}_{n}) $. This dataset is only used for training purpose. In the prediction phase, the collected PDP data is used instead as input to the trained model for trajectory estimation of the user terminal.

\section{Attention-Based Mechanism for BFF Trajectory Estimation}
\subsection{Utilization of BFFs for Trajectory Modeling}
The previous section described the method by which we can convert a single BFF into a position. After extracting the positions in the environment based on BFFs, trajectories can be created. The dataset did not include any segmentation of movement types or traffic rules. In a real-world dataset, these aspects would likely be observed, which could further enhance the accuracy of the estimator. 

Each trajectory can be described as $\textbf{Y} = [\textbf{y}^{(1)}, \textbf{y}^{(2)},\cdots ,\textbf{y}^{(L)}] \in \mathbb{R}^{1\times 2L}$, where $L$ is the trajectory length (or the size of the system's memory), and each $\textbf{y}$ shows a 2D location in the environment which is describes in the previous section by $\textbf{y}_{n}$. We assume that all paths or trajectories are involved in a movement in the environment. The input of our models consists of the trajectories of different users, represented as $\textbf{Y}^{\prime} = [ \textbf{Y}^{}_{1},\textbf{Y}^{}_{2},\cdots ,\textbf{Y}^{}_{N}]^{\text{T}}$, where $\textbf{Y}^{\prime}\in \mathbb{R}^{ N \times 2L}$ and row space of the above matrix is considered as a 2D cartesian path. We observe the positions of all paths from time step 1 to $\text{T}_{obs}$, and predict their positions for the time step from $\text{T}_{obs}+1$ to $L$. The set of future trajectories can be described as $\bar{\textbf{Y}} \in \mathbb{R}^{N \times 2 }$ and our estimated trajectories through the models can be denoted as $\hat{\textbf{Y}} \in \mathbb{R}^{N \times 2 }$.
 
\subsection{Trajectory Estimation Using TNs}
TNs have been successfully used to estimate datasets that are structured as sequences. One of the main benefits of using TN is their ability to focus on subtle details and recognize different patterns within long data sequences.
In addition, TN utilizes an encoder-decoder design to accurately map inputs to the corresponding outputs.

\begin{figure}
\centering
\includegraphics[width=0.4\textwidth]{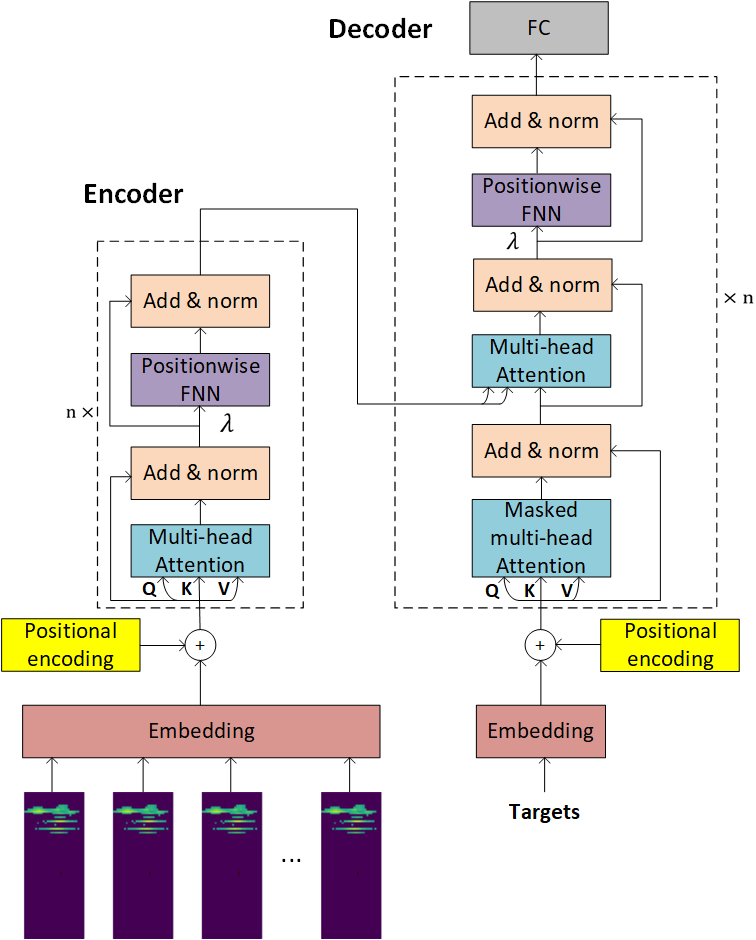}
\caption{\label{fig:frog} Representation of a TN model applied to BFF data.}
\label{transformer}
\end{figure}

The encoder-decoder structures are widely utilized in sequence-to-sequence models. In this paper, the encoder block transforms the input sequence $\textbf{Y}^{\prime}$ into a latent state $\textbf{Z}^{} = \left[ \textbf{Z}_{1}^{}, \textbf{Z}_{2}^{},\cdots ,\textbf{Z}_{N}^{} \right]$. Given $\textbf{Z}^{}$, $\hat{\textbf{Y}}$ is produced at each time step. Note that, in an autoregressive manner, the previously generated element is used as an additional input to generate the subsequent element. In the TN, both the encoder and decoder blocks use stacked self-attention, position-wise, and fully connected (FC) layers. These layers can be seen in the left and right halves of Fig. \ref{transformer}, respectively.  For an in-depth discussion of the internal structures of TNs, we refer the reader to  \cite{he2016deep}.

The TN mainly can capture the dependencies in sequential data and the non-linear attributes of BFFs through its Attention module. This module is composed of a self-attention mechanism and a multi-head attention sub-layer. These sub-layers allow the model to focus on subtle details of the input sequence, capturing various aspects of the information.

The self-attention takes these inputs: the query matrix $\textbf{Q}$, the key matrix $\textbf{K}$, the value matrix $\textbf{V}$, and the dimension $d_k$ of the queries and keys. It computes the dot product of $\textbf{Q}$ and $\textbf{K}$, and then scales it by a factor of $\frac{1}{\sqrt{d_k}}$. The weights of values are determined by applying a softmax function. This scaled operation is particularly beneficial when $d_k$ is large, as it prevents the softmax function from entering regions where the gradients are extremely small.

\begin{equation}   \label{eq4}
\text{Attention} (\textbf{Q}, \textbf{K}, \textbf{V}) = \text{softmax}(\frac{\textbf{Q}\textbf{K}^{\text{T}}}{\sqrt{d_k}})\textbf{V}.
\end{equation}

Rather than learning a single attention function, it has been found advantageous to map the queries, keys, and values $h$ times in order to learn different contextual information for each \cite{vaswani2017attention}. The self-attention function applies to each projected version of the queries, keys, and values. Following this, the outputs are concatenated and projected once more to determine the best and final value weights. Thus, by leveraging the multi-head attention sub-layer, the TN is capable of simultaneously producing a comprehensive latent feature of the trajectory data.
\begin{equation}   \label{eq5}
\begin{split}
\text{MultiHead}(\textbf{Q}, \textbf{K}, \textbf{V}) = \text{concat}(\text{head}_1,\cdots, \text{head}_h)\boldsymbol{\textbf{W}}^{O},\\
\text{head}_{i} = \text{Attention}(\textbf{Q}\boldsymbol{\textbf{W}}_{i}^{Q}, \textbf{K}\boldsymbol{\textbf{W}}_{i}^{K}, \textbf{V}\boldsymbol{\textbf{W}}_{i}^{V}  ).
\end{split}
\end{equation}

The parameters matrices in queries, keys, values, and output are represented by the projections $\boldsymbol{\textbf{W}}_{i}^{Q}, \boldsymbol{\textbf{W}}_{i}^{K}, \boldsymbol{\textbf{W}}_{i}^{V}$ and $\boldsymbol{\textbf{W}}_{i}^{O}$, respectively.

Moreover, the fully connected Feed-Forward Networks are made up of two linear transformations, with a ReLU activation function that is applied to each of the attention sublayers.
\begin{equation}
    \text{Feedforward}(\boldsymbol{\lambda}) = \text{max}(0, \boldsymbol{\lambda}\boldsymbol{\textbf{W}}_{1}+\textbf{b}_{1})\boldsymbol{\textbf{W}}_{2} + \textbf{b}_{2}.
\end{equation}

Similar to other sequence-based models, this particular model uses learned embeddings to convert input and output tokens into vectors with a dimension of $\text{T}_{obs}$. A learned linear transformation function is also applied to the decoder output to generate predicted probabilities for the next token. The two embedding layers share a common weight matrix. In addition, positional encoding is a technique used in the TN to help the model capture the sequential order of the input data. Unlike RNNs, Transformers do not have a natural way of encoding the order of the sequence, because the model treats all the input tokens as unordered elements. Therefore, Positional Encoding provides a way to inject information about the sequence position into the input embeddings. Positional Encoding works by adding a fixed vector to the embedding vector of each input token, which encodes the token's position within the sequence. The added vector is calculated based on a set of sinusoidal functions with different frequencies and phases \cite{gehring2017convolutional}. Specifically, the Positional Encoding for a token at position $pos$ and dimension $j$ is given by:
\begin{equation}   \label{eq7}
\text{PE}_{{(pos, j)}} = 
\begin{cases}
  \sin({pos}/10000^{2j/\text{T}_{obs}}) & \text{if $j$ is even} \\
  \cos({pos}/10000^{2(j-1)/\text{T}_{obs}}) & \text{if $j$ is odd}
\end{cases}.
\end{equation}

\section{Numerical Results}
We conducted experiments to evaluate the trajectory estimation performance of various DL architectures. Initially, we learned the trajectories using trained deep neural networks based on BFF data. Following this, we introduced new, unseen trajectories that are not part of the training set for the testing stage. The performance of these networks is then evaluated. Ultimately, we compared the performance of our proposed method with other state-of-the-art DL methods, specifically looking at average error versus sequence length. The parameters for the simulation settings are detailed in Table \ref{t1}. To provide a clearer understanding of the results, we have divided them into two subsections: Simulation Methodology and Results and Discussion.

\begin{table}[]
\caption{The DL architectures hyperparameters.\label{long}}
\centering
\begin{tabular}{ |c|c|c| }
\cline{1-2}
\textbf{Parameters}              & \textbf{All tracking-based approaches} \\
\cline{1-2}
Number of units         &   (256, 64)    \\
\cline{1-2}
MLP layers              &   2    \\
\cline{1-2}
Regression output       &   2 neurons (2D position)\\
\cline{1-2}
\#of training sequences &    320408\\
\cline{1-2}
Epochs                  &   Up to 100
\\
\cline{1-2}
Batch size              &   64 \\
\cline{1-2}
Optimizer               &   ADAM 
\\
\cline{1-2}
Learning rate           &   $10^{-3}$   \\
\cline{1-2}
Dropout                 &      0.01  \\
\cline{1-2}
\end{tabular}
\label{t1}
\end{table}
 
\subsection{Simulation Methodology}

We utilize the Wireless InSite ray-tracing simulator to generate mmWave data \cite{Remcom}. This allows us to analyze the propagation of wireless signals in complex urban environments and assess the accuracy of our model. We examine a detailed 3D map of the New York University (NYU) region \cite{nework}, which includes BFF data from 160801 distinct two-dimensional locations (401 x 401 positions). In accordance with the findings in \cite{azar201328}, we employ the propagation parameters and the ray-tracing simulations, which are found to be consistent with experimental measurements \cite{gante2018data}. To evaluate the performance of our beamforming algorithm, we conduct simulations at a carrier frequency of $28$GHz with a codebook size of $M = 32$. For all potential transmit beamforming vectors, we set the maximum received power to be $30$dBm. The transmitter is positioned at coordinates $(0,0)$, and we add noise into the ray-tracing data using a log-normal distribution with a standard deviation of $6$dB.

\begin{figure}
\centering
\includegraphics[width=0.46\textwidth]{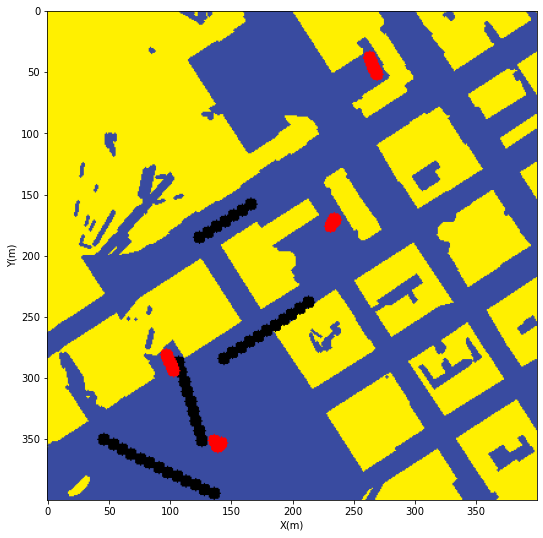}
\caption{\label{fig:frog}  
The generated sequence in the NYU area}
\label{NYU}
\end{figure}

To create sequences for tracking-based methods, two types of synthetic sequences are produced: pedestrian-like and vehicle-like sequences. The pedestrian-like sequences are designed to allow sudden stops or quick changes in direction at a low average speed of 5 km/h, while the vehicle-like sequences have a higher average speed of 30 km/h and acceleration, with a limited steering angle. The sequences in the dataset are sampled at a rate of 1 Hz, producing paths as shown in Fig. \ref{NYU}. With regard to sequences, the training, validating, and testing paths are selected from separate groups of trajectories in order to avoid memorization.

\subsection{Results and Discussion}
We utilized three DL models commonly used for tracking and trajectory estimation for benchmarking against our proposed method. These include LSTM, temporal convolutional network (TCN) \cite{gante2020deep}, and RNN. Table \ref{t2} presents a comparison of three significant attributes: the model size in terms of learnable parameters, training time, and prediction time. As observed, TNs is faster than LSTM and TCN models, as this desired method consider all inputs simultaneously. Training LSTM and TCN networks poses a greater challenge than TNs due to their significantly larger number of learnable parameters. Despite this, TNs maintain competitive complexity and computational cost with RNN.


\begin{table}[]
\caption{Number of learnable parameters, training and prediction  time for sequence length 11 for the vehicles.\label{long}}
\centering
\begin{tabular}{ |c|c|c|c| }
\cline{1-4}
\textbf{DL architecture}     & \textbf{L.parameters} & \textbf{T.time (hours)} & \textbf{P.Time (mins)} \\

\cline{1-4}
RNN          & $6.525\times10 ^{4}$           & $00:39:51$                   & $00:45$                      \\
\cline{1-4}
LSTM                & $2.625\times10 ^{5}$           & $2:41:50$                   & $05:17$                     \\
\cline{1-4}
TCN                & $2.812\times10 ^{5}$           & $2:31:03$                   & $04:25$ \\
\cline{1-4}
Transformer & $6.9\times10 ^{4}$           & $00:53:15$                   & $01:07$ \\
\cline{1-4}
\end{tabular}
\label{t2}
\end{table}

\begin{figure}
\centering
\includegraphics[width=0.48\textwidth]{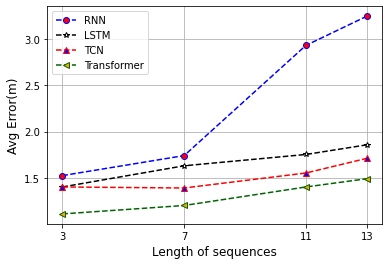}
\caption{\label{fig:frog} Prediction error comparison 
 vs sequence length 
for vehicles, considering an average noise value ($\sigma$ = 6 dB).}
\label{vehicles}
\end{figure}

Fig. \ref{vehicles} illustrates the average error achieved by the DL models between $\bar{\textbf{Y}} \in \mathbb{R}^{N \times 2 }$ and $\hat{\textbf{Y}} \in \mathbb{R}^{N \times 2 }$ for the trajectory estimation task of vehicles across different BFF sequences. It is evident that in  our proposed method outperforms the state-of-the-art methods in terms of average error, maintaining precision even for longer trajectories. In addition, vehicle-like sequences present distinctive characteristics, including higher average speeds and rapid accelerations. Consequently, deep learning-based trajectory estimation for these sequences tends to be slightly less precise than for sequences with less complexity in movement. The increased complexity of vehicle-like motions poses challenges for accurate modeling, leading to some compromises in the performance of trajectory estimation.

\begin{figure}
\centering
\includegraphics[width=0.48\textwidth]{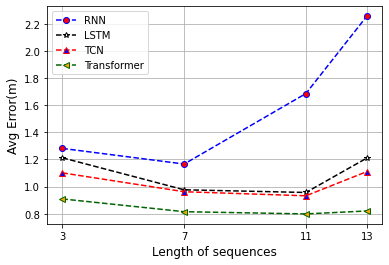}
\caption{\label{fig:frog}Prediction error comparison 
vs sequence length for pedestrians, considering an average noise value ($\sigma$ = 6 dB).}
\label{ped}
\end{figure}

Fig. \ref{ped} further depicts the average error versus sequence length for pedestrians using the same DL methods. As observed, our proposed method continues to surpass other methods. The inherent predictability and smoothness of pedestrian-like sequences make them ideal for accurate modeling and tracking.

\begin{figure}
\centering
\includegraphics[width=0.48\textwidth]{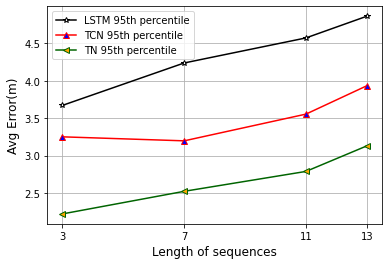}
\caption{\label{fig:frog}Comparative analysis of $95^{th}$ percentile errors 
vs sequence length for vehicles, considering an average noise value ($\sigma$ = 6 dB).}
\label{95vehicle}
\end{figure}

\begin{figure}
\centering
\includegraphics[width=0.48\textwidth]{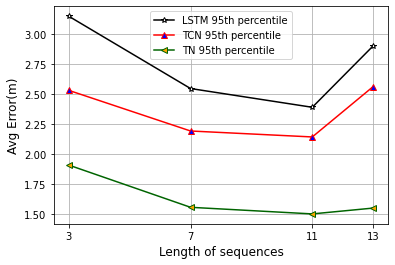}
\caption{\label{fig:frog}Comparative analysis of $95^{th}$ percentile errors
vs sequence length for pedestrians, considering an average noise value ($\sigma$ = 6 dB).}
\label{95ped}
\end{figure}

Fig. \ref{95vehicle} and Fig. \ref{95ped} demonstrate $95^{th}$ percentile error as a tool to extract additional accuracy across the proposed approach TN, TCN, and LSTM at noise level 6 dB for vehicles and pedestrians, respectively. As can be seen, the TN still outperforms in terms of $95^{th}$ percentile errors than the two other state-of-the-art methods. The $95^{th}$ percentile basically says that 95 percent of the error is below this value and the other 5 percent of the error exceeds that value. Moreover, we compare the performance of the models for a sequence length = 7 and noise level $\sigma=9$dB in Table \ref{t3}. The table demonstrates the superior performance of the TN model under these conditions. Despite the increased noise level, the TN model consistently outperformed both the LSTM and TCN models.


\begin{table}[h]
\caption{Comparing the performance of the models for a sequence length = 7 and noise level $\sigma=9$ dB.\label{long}}
\centering
\begin{tabular}{ |c|c|c| }
\cline{1-3}
\textbf{DL architecture}     & \textbf{Vehicles Avg Error(m)} & \textbf{Pedestrians Avg Error(m)}\\

\cline{1-3}
LSTM          & $2.20712$           & $1.0045$\\
\cline{1-3}
TCN                & $1.69597$           & $0.982$\\
\cline{1-3}
Transformer               & $1.4104$           & $0.89$\\
\cline{1-3}
\end{tabular}
\label{t3}
\end{table}

\section{Conslusion}
In this paper, we proposed and investigated a procedure for trajectory estimation in mmWave MIMO outdoor systems. We show that TN architecture can capture sequences of BFF signals originating from vehicles and pedestrians' movements, and through the application of learning mechanisms, subsequently estimate these trajectories. Our simulation results confirm that the TN architecture based on the attention mechanism outperforms the other state-of-the-art methods.

\bibliographystyle{IEEEtran}
\bibliography{sample.bib}

\end{document}